# Electronic Doping and Scattering by Transition Metals on Graphene

K. Pi,* K. M. McCreary,* W. Bao, Wei Han, Y. F. Chiang, Yan Li, S.-W. Tsai, C. N. Lau, and R. K. Kawakami

**Department of Physics and Astronomy, University of California, Riverside, CA 92521**

**e-mail: roland.kawakami@ucr.edu**

**Abstract:**

We investigate the effects of transition metals (TM) on the electronic doping and scattering in graphene using molecular beam epitaxy combined with *in situ* transport measurements. The room temperature deposition of TM onto graphene produces clusters that dope *n-type* for all TM investigated (Ti, Fe, Pt). We also find that the scattering by TM clusters exhibits different behavior compared to 1/*r* Coulomb scattering. At high coverage, Pt films are able to produce doping that is either *n-type* or weakly *p-type*, which provides experimental evidence for a strong interfacial dipole favoring *n-type* doping as predicted theoretically.

PACS numbers: 73.63.-b, 72.10.Fk, 73.40.Ns, 73.23.-b

*These authors contributed equally to the work.

# I. INTRODUCTION

Transition metal (TM) adatoms and clusters on graphene have recently been a topic of great interest: at low density, they are expected to induce doping, scattering [1], and novel magnetic [2-4] and superconducting [5] behavior; at high density (up to continuous coverage), they may locally dope or modify the band structure of graphene [6-8]. Because of their importance for graphene-based electronics and the investigation of novel phenomena [1-14], there have been extensive theoretical studies [1-8, 13, 14]. In contrast, the experimental exploration of TM/graphene systems is much more limited.

A key issue to investigate is the charge transfer between the TM and graphene because it is responsible for both the local doping and the charge impurity scattering. Generally, the relative work function (WF) between the TM and the graphene are believed to be important factors for determining the charge transfer [11], *i.e.* graphene will be *p*-doped (*n*-doped) if the TM's WF is larger (smaller) than graphene. Recently, density functional calculations predict the presence of a strong interfacial dipole that promotes the *n-type* doping of graphene [8]. However, experimental studies of the local doping by TM contacts have yet to find evidence for this strong interfacial dipole layer [9-12].

In this work, we report *in situ* transport measurements of TM/graphene systems as a function of TM coverage for several different metals, using a molecular beam epitaxy (MBE) deposition system with built-in electrical probes. Similar techniques have been applied to study gases [15], molecules [16], and alkali metal [17] adatoms on graphene. The metals used in the study are Ti, Fe, and Pt, with WF of 4.3, 4.7, and 5.9 eV [18], respectively (the WF of graphene is 4.5 eV [19, 20]). Surprisingly, at low coverage, the TM clusters dope graphene *n-type*, regardless of its WF relative to that of graphene. For the scattering at low coverage, we find that the scattering by TM clusters exhibits different behavior compared to $1/r$ Coulomb scattering. Extending to high coverage, we make the important observation that Pt films are able to produce doping that is either *n-type* or weakly *p-type*. Because WF considerations alone would predict strong *p-type* doping, this result provides experimental evidence for the strong interfacial dipole favoring *n-type* doping as calculated theoretically [8].

## II. EXPERIMENTAL PROCEDURES

Samples are prepared by mechanical exfoliation of Kish graphite onto a $SiO_2$/Si substrate (300 nm thickness of $SiO_2$). Single layer graphene flakes are identified by optical microscopy and Raman spectroscopy [21]. Figure 1a shows a scanning electron microscope (SEM) image of a typical graphene device with Au/Ti electrodes defined by e-beam lithography. The device is annealed under Ar/$H_2$ environment at 200°C for one hour to remove resist residue [22, 23] and degassed in ultrahigh vacuum at 90°C for one hour. The room temperature MBE deposition of TM atoms (growth pressure $< 7\times10^{-10}$ torr) is calibrated by a quartz deposition monitor. The coverage is converted from atoms/cm$^2$ to "monolayers" (ML) where 1 ML is defined as $1.908\times10^{15}$ atoms/cm$^2$, the areal density of primitive unit cells in graphene. For low coverage, the room temperature deposition of TM leads to clustering as shown in the atomic force microscope (AFM) image of 0.01 ML Pt on graphene (Figure 1b). The presence of isolated adatoms cannot be ruled out by the AFM, but are unfavorable theoretically [6]. *In situ* transport measurements are performed using standard lock-in detection (1 μA excitation).

## III. RESULTS AND DISCUSSION

The fine control of TM deposition provides the ability to probe the effect of small amounts of material on the transport properties of graphene. Figure 1c shows representative gate dependent conductivity scans for various thicknesses of Ti in the low coverage regime. The minimum in the gate dependent conductivity identifies the position of the Dirac point ($V_D$), while the slope corresponds to the mobility of charge carriers in the graphene. With increasing coverage, two characteristic behaviors are observed. First, the introduction of Ti on the graphene surface results in shifting the Dirac point towards more negative gate voltages, indicating that the Ti is a donor, producing *n-type* doping in the graphene. Second, the slope of the conductance curves away from the Dirac point decreases, indicating that the Ti introduces additional scattering to lower the mobility. Both of these characteristics are also observed for Fe doping (Figure 1d) and Pt doping (Figure 1e).

Figure 2 highlights the relation between the Dirac point shift ($V_{D,shift} = V_D - V_{D,initial}$) and TM coverage for a collection of Ti, Fe, and Pt samples in the low coverage regime. Despite the sample-to-sample variations which may be due to differences in the graphene surface purity, growth rate uncertainties, and the possible dependence of graphene WF on flake size or edge roughness [24], several important features are discovered. First, all samples, including the Pt samples with WF greater than graphene, result in *n-type* doping. Second, the three different TM result in three different ranges for slopes, with the Ti samples exhibiting the most negative initial slopes (-2169 to -4602 V/ML). From this value the doping efficiency, or number of electrons transferred per Ti atom to graphene is determined by knowing the carrier concentration associated with the given change in gate voltage ($\Delta n = \alpha \Delta V_g$, where $\alpha = 7.2\times10^{10}$ $V^{-1}cm^{-2}$ based on calculated capacitance values). The doping efficiency is in the range of 0.082 to 0.174 electrons per Ti atom. The Fe shows the next strongest efficiency (0.017 to 0.046), while the Pt is the weakest electron donor with the efficiency of 0.014 to 0.021 electrons transferred for each Pt atom. Upon recalling the bulk WFs of Ti (4.3 eV), Fe (4.7 eV), and Pt (5.9 eV), it is apparent that the WF of the TM is related to the doping efficiency, with electrons being more easily transferred from the lowest WF material, Ti, compared to the highest WF material, Pt. However, the magnitudes of the doping efficiency do not vary linearly with the WF of the TM. Therefore, in addition to the work function, other effects such as wave function hybridization or structural modifications may contribute to the electronic doping of graphene.

Figures 3a-3c show the conductivity as a function of carrier concentration [$n = -\alpha(V_g - V_D)$]. The electron and hole mobilities are determined by taking the slope of the conductivity away from the Dirac point ($\mu_{e,h} = |\Delta\sigma/e\Delta n|$) [15, 17]. Figures 3d-3f illustrate the detailed dependence of mobility on the TM coverage for Ti, Fe, and Pt samples in the low coverage regime. Comparing the different samples at equivalent coverages, the Ti exhibits the strongest scattering and Pt has the weakest scattering. Noting that the trend in the scattering (Ti > Pt) matches that of the doping efficiency, we investigate this relationship by plotting the normalized mobility [25] against the Dirac point shift (Figure 3g). The average mobility, $\mu = (\mu_e + \mu_h)/2$, is plotted for Ti and Pt. The Fe samples typically exhibit a reduction of hole mobility which is most pronounced in sample Fe-2, so $\mu_e$ and $\mu_h$ are plotted separately. Comparing the different materials shows that the mobility reduction of Ti, Pt, and Fe (electrons) is much more strongly related to the Dirac point shift than the TM coverage (Figure 3g). Because the Dirac point shift not only measures the doping level in the graphene but also the average charge density of the TM, the data shows that the scattering is related to the average charge density of the clusters—a characteristic that is plausible for Coulomb scattering. However, we point out that this behavior is actually different from what is calculated for Coulomb scattering by point-like scatterers with $1/r$ potential [26]. Specifically, in ref. 26, the scattering per impurity does not scale linearly with the impurity charge ($\alpha_\varepsilon$) and instead has a strong quadratic component, resulting in scattering that scales as $\alpha_\varepsilon^2 n_{imp} = \alpha_\varepsilon(\alpha_\varepsilon n_{imp}) \sim \alpha_\varepsilon(V_{D,shift})$. Due to the presence of the material-dependent $\alpha_\varepsilon$ factor (i.e.

doping efficiency), the mobility vs. Dirac point shift curves should be significantly different for different materials. Therefore, the observed scattering by TM clusters exhibits behavior that differs from $1/r$ Coulomb scattering by isolated impurities [1].

Additionally, we analyze the power law relationship between the scattering and doping effects. The total scattering rate is $\Gamma = \Gamma_0 + \Gamma_{TM}$, where $\Gamma_0$ is the scattering rate of the undoped sample and $\Gamma_{TM}$ is the scattering rate induced by the TM. Because mobility is inversely proportional to scattering, the quantity $1/\mu - 1/\mu_0$ is proportional to $\Gamma_{TM}$. The relationship between the Dirac point shift and $\Gamma_{TM}$ is investigated by plotting $-\Delta V_{D,shift}$ vs. $1/\mu - 1/\mu_0$ on a log-log scale (Figure 3h). The dashed lines are power law fits, $-\Delta V_{D,shift} \sim (\Gamma_{TM})^b$, with values of $b$ ranging from 0.64 – 1.01 as indicated in the figure caption. Compared to the results of Chen et. al. [17] which find values of $b$ = 1.2-1.3 for scattering by isolated potassium impurities, our results with $b \leq 1$ indicate a different behavior for scattering by TM clusters.

A surprising result from the studies at low coverage (Figures 1-3) is the *n-type* doping of graphene by Pt. If the WF is the only factor affecting the transfer of electrons between materials, Pt is expected to dope graphene strongly *p-type*, since the WF of Pt (5.9 eV) is significantly larger than that of graphene (4.5 eV). Density functional calculations of bulk TM on graphene [8] present a possible explanation for this observed behavior by predicting the formation of an interfacial dipole layer, resulting in a potential step to favor *n-type* doping ($\Delta V$ = 0.9 eV). So far, however, there has been no experimental evidence for such a strong dipole layer forming at the interface between a bulk TM and graphene [9-12]. To investigate the theoretical prediction of a strong interfacial dipole layer between the graphene and bulk TM, we extend the Pt-doping study to higher coverage to study the charge transfer from Pt films. Figure 4a displays $V_D$ as a function of coverage for several Pt-doped samples. An initial rapid shift toward negative voltages is observed in all samples. As more Pt is deposited, bringing the sample into the medium coverage regime, the rate of shift in $V_D$ slows and reaches a minimum value before gradually increasing towards more positive voltages. At high coverage, the Dirac point stabilizes and shows very little variation with additional deposition. The sample morphology is measured by *ex situ* AFM. The AFM image for 0.62 ML of Pt shows that the Pt is still in the form of isolated clusters (Figure 4b). At the higher coverage of 3.19 ML, the Pt forms a connected film with some uncovered regions of graphene (Figure 4c). The connected film provides a parallel conduction pathway that contributes to the measured conductivity value, but should not be gate dependent. The gate dependence of the conductivity is primarily due to the chemical potential shift of the graphene that is not covered by the metal. For graphene in direct contact with the metal, the local chemical potential is pinned, exhibiting no gate dependence. However, the gate dependence of the uncovered graphene regions and the voltage of the conductance minima ($V_D$) still provide a reliable measure of the electronic doping by the TM due to the continuity of the chemical potential. Thus, the final values of $V_D$ in the high coverage regime clearly show that Pt films can produce either *n-type* or weak *p-type* doping of the graphene. This sample-to-sample variation is most likely due to differences in the initial surface purity among samples. Although hydrogen cleaning is performed on all samples, trace amounts of resist residue could remain, directly affecting the TM-graphene spacing. Due to the highly spacing-dependent interfacial dipole strength,[8] any variation in the spacing will directly affect the type and amount of doping. The fact that *n-type* doping is observed provides experimental evidence for the presence of a strong interfacial dipole layer favoring *n-type* doping as predicted theoretically [8] because the expected doping based only on WF considerations would lead to strong *p-type* doping.

An interfacial dipole whose strength decreases with increasing equilibrium spacing ($d_{eq}$) [8] provides a possible explanation for the non-monotonic behavior of the Dirac point shift in Pt samples. Based on theoretical calculations, the $d_{eq}$ between TM adatoms and graphene is less than 3 Å [6] while for bulk TM the distance increases to ~3.3 Å [8]. The *n-type* doping observed in samples at low coverage is an indication of a strong interfacial dipole favoring *n-type* doping, as expected for low coverages exhibiting a small $d_{eq}$. As the bulk-like regime is approached, the increasing $d_{eq}$ decreases the dipole strength and hence reduces the *n-type* doping efficiency as observed by the shift in the Dirac point toward positive voltages. We emphasize that the interfacial dipole provides just one possible scenario to explain the non-monotonic evolution of the Dirac point shift. A quantitative understanding is complicated by the fact that the WF can differ from bulk values for small clusters (< 4 nm lateral size) [27] and the corresponding quantity for adatoms (should they be present) is the first ionization energy. Therefore, further theoretical calculations are needed to fully understand the doping effect of clusters. Regardless of the exact mechanism for doping by clusters, an interfacial dipole is still necessary to explain the *n-type* or weak *p-type* doping measured in the bulk-like regime.

## IV. CONCLUSION

In conclusion, the exploration of TM/graphene systems leads to several important observations. At low coverage, the doping efficiency is found to be related to the TM WFs, but Ti, Fe, and Pt all exhibit *n-type* doping even for materials with higher WF than graphene (i.e. Fe, Pt). Extending the Pt doping study to higher thickness, the doping can either be *n-type* or weakly *p-type*. Because WF considerations alone would generate strong *p-type* doping, this result provides experimental evidence for the strong interfacial dipole favoring *n-type* doping as predicted by theory [8]. Analysis of the scattering at low coverage indicates that the scattering by TM clusters exhibits different behavior compared to 1/*r* Coulomb scattering.

V. ACKNOWLEDGEMENTS

We acknowledge helpful discussions with F. Guinea and M. Fuhrer. We acknowledge the support of ONR (N00014-09-1-0117), NSF (CAREER DMR-0450037), NSF (CAREER DMR-0748910), NSF (MRSEC DMR-0820414), and CNID (ONR/DMEA-H94003-07-2-0703).

FIGURE CAPTIONS

Figure 1. (a) SEM image of a graphene device with Au(100 nm)/Ti(10 nm) electrodes. (b) AFM image of 0.01 ML of Pt deposited on single layer graphene. (c-e) The gate dependent conductivity at selected TM coverage for Ti, Fe, and Pt, respectively.

Figure 2. Dirac point shift vs. coverage for nine separate samples. The dashed lines indicate the linear fit used to define the doping efficiencies, which are: 0.174, 0.092, and 0.082 electrons/atom for Ti-1, Ti-2 and Ti-3, respectively; 0.017, 0.040 and 0.046 electrons/atom for Fe-1, Fe-2 and Fe-3, respectively; 0.014, 0.021, and 0.019 electrons/atom for Pt-1, Pt-2 and Pt-3 respectively.

Figure 3. (a-c) The conductivity vs. carrier concentration for Ti, Fe, and Pt, respectively, for four different TM coverages. (d-f) The electron and hole mobilities for Ti, Fe, and Pt, respectively, as a function of TM coverage. (g) The normalized mobility ($\mu/\mu_0$) vs. Dirac point shift. The data points corresponding to 0.102 ML Pt, 0.008 ML Ti and 0.029 ML Fe on samples Pt-1, Ti-1 and Fe-2 are circled. (h) $-V_{D,shift}$ is plotted vs. $1/\mu - 1/\mu_0$. The dashed lines are power law fits to the equation, $-\Delta V_{D,shift} \sim (\Gamma_{TM})^b$ where $b$ is 0.64, 1.01, 0.85, 0.83, 0.86 and 0.95 for Ti-1, Ti-2, Pt-1, Pt-2, Pt-3 and Fe-2 electrons, respectively.

Figure 4. (a) The Dirac point as a function of Pt coverage up to high coverage. (b) AFM image of 0.62 ML Pt exhibits isolated clusters. (c) AFM image of 3.19 ML Pt indicates a connected film with some areas of bare graphene.

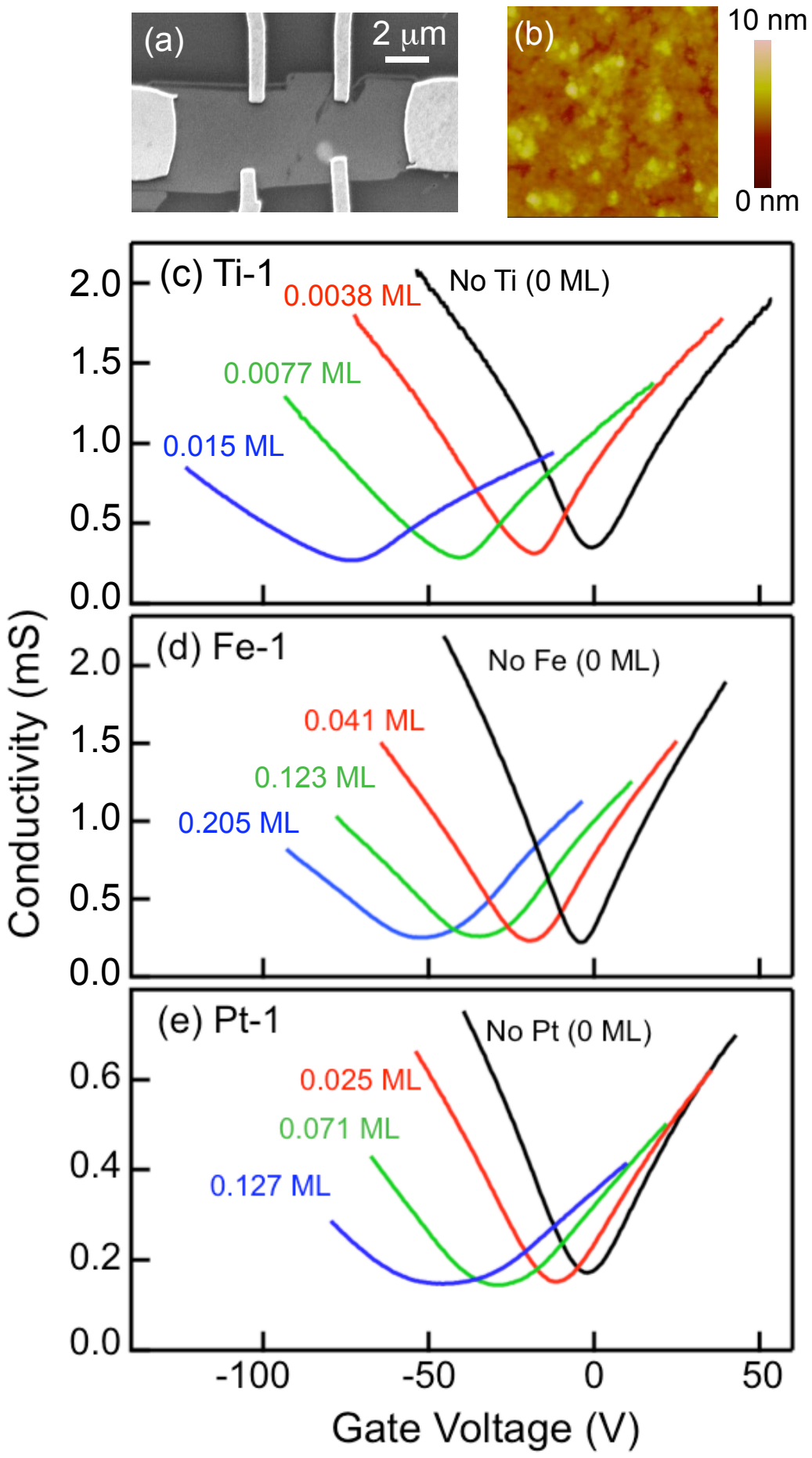

Figure 1.

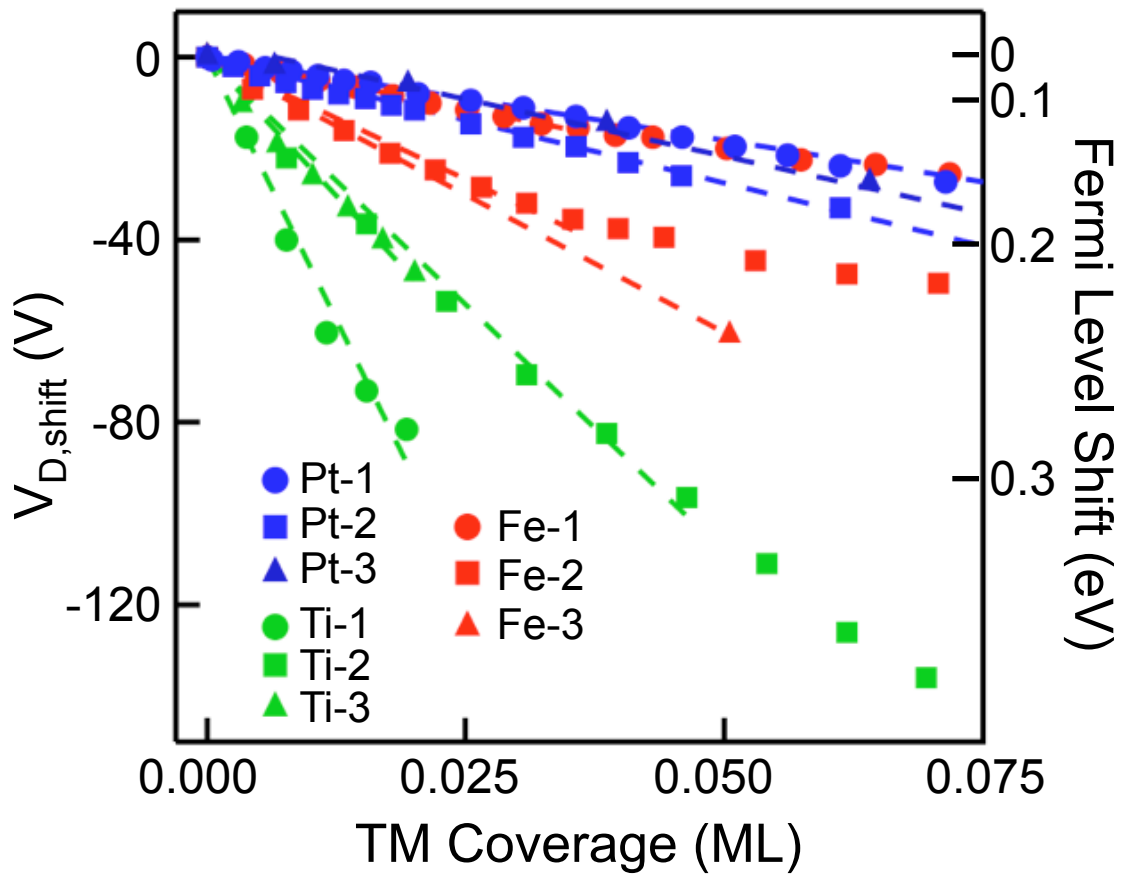

Figure 2.

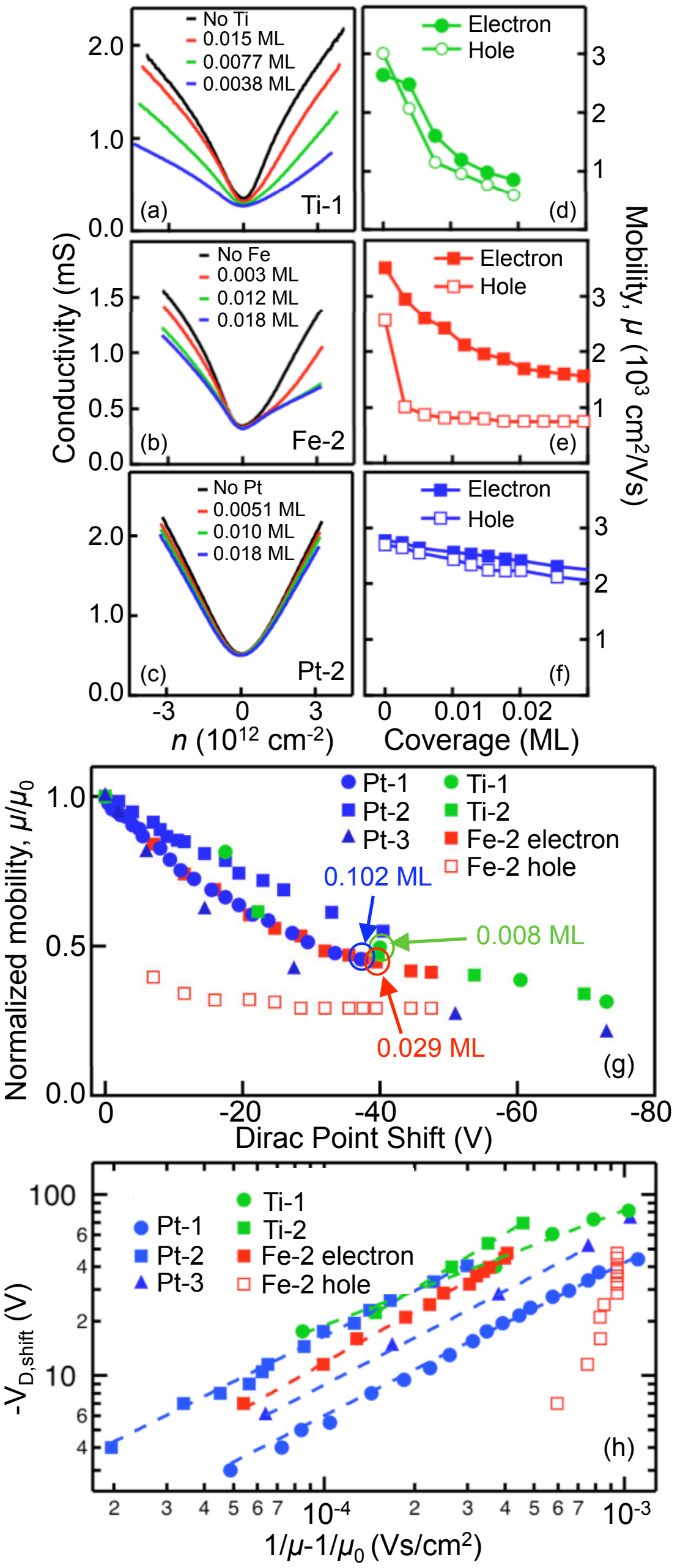

Figure 3.

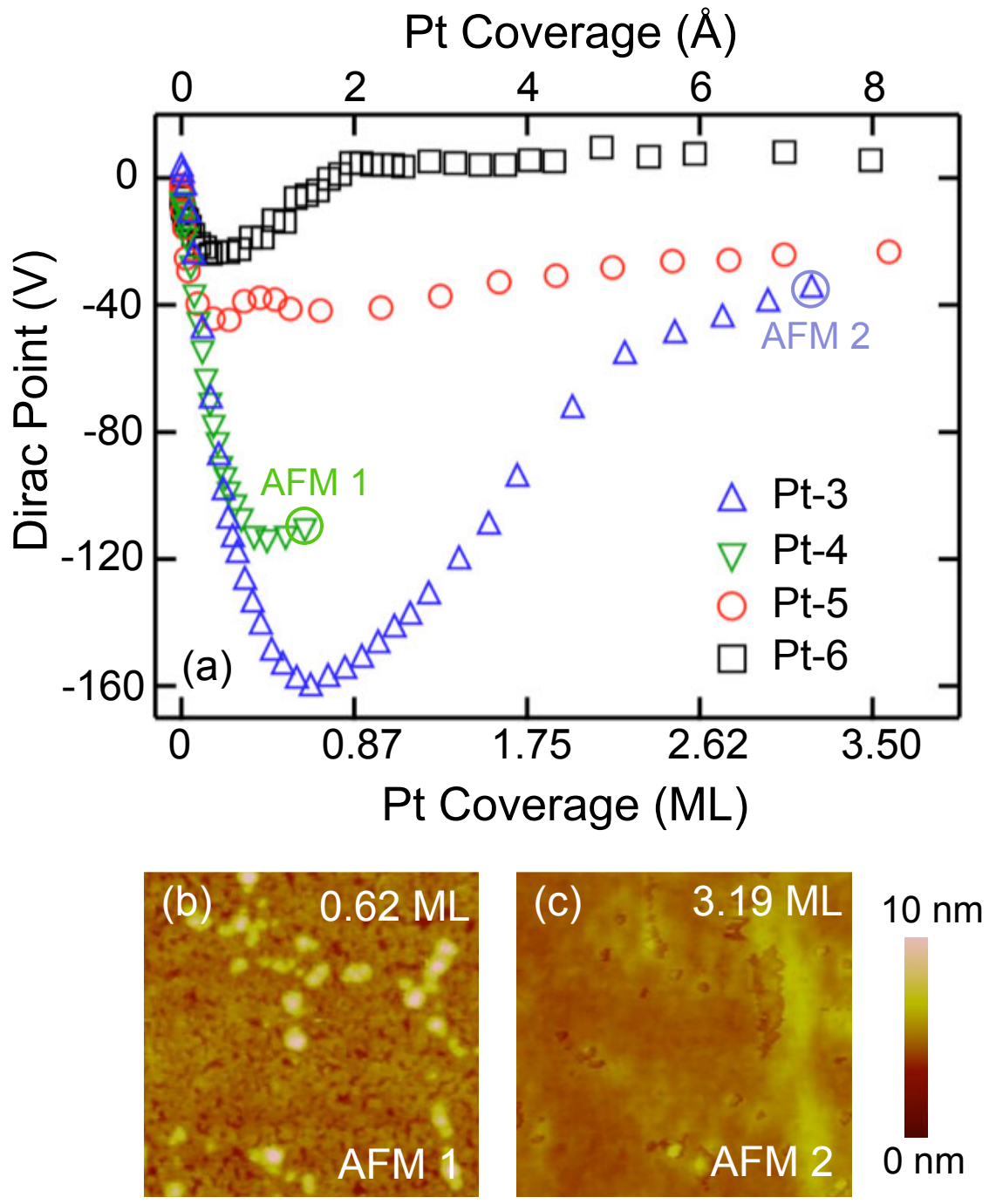

Figure 4.